\documentclass[preprint]{aastex}

\newcommand{\HST}{{\sl HST}}

\newcommand{\Lsun}{\mbox{$L_{\sun}$}}

\newcommand{\Mjup}{\mbox{$M_{Jup}$}}

\newcommand{\perone}{\mbox{$^{-1}$}}

\newcommand{\etal}{et al.}
\newcommand{\eg}{e.g.}
\newcommand{\ie}{i.e.}

\newcommand{\Kp}{\mbox{$K^{\prime}$}}

\newcommand{\degs}{\mbox{$^{\circ}$}}

\newcommand{\Teff}{\mbox{$T_{\rm eff}$}}

\newcommand{\eInd}{\hbox{$\epsilon$~Ind}}
\newcommand{\sdssbin}{\hbox{SDSS~J1534+1615}}

\slugcomment{ApJ, accepted April 28, 2006}

\shorttitle{A Novel T Dwarf Binary from LGS AO}
\shortauthors{Liu \etal}

\begin{document}

\title{SDSS~J1534+1615AB: A Novel T Dwarf Binary Found with
  Keck Laser Guide Star Adaptive Optics and the Potential Role of Binarity in
  the L/T Transition\altaffilmark{1}}


\author{Michael C. Liu,\altaffilmark{2,3}
        S. K. Leggett,\altaffilmark{4}
        David A. Golimowski,\altaffilmark{5}\\
        Kuenley Chiu,\altaffilmark{5}
        Xiaohui Fan,\altaffilmark{6}
        T. R. Geballe,\altaffilmark{7}
        Donald P. Schneider,\altaffilmark{8}
        J. Brinkmann\altaffilmark{9}}

\altaffiltext{1}{The data presented herein were obtained at the
  W.M. Keck Observatory, which is operated as a scientific partnership
  among the California Institute of Technology, the University of
  California, and the National Aeronautics and Space Administration. The
  Observatory was made possible by the generous financial support of the
  W.M. Keck Foundation.}
\altaffiltext{2}{Institute for Astronomy, University of Hawai`i, 2680
  Woodlawn Drive, Honolulu, HI 96822; mliu@ifa.hawaii.edu}
\altaffiltext{3}{Alfred P. Sloan Research Fellow}
\altaffiltext{4}{United Kingdom Infrared Telescope, Joint Astronomy
  Centre, 660 North A'ohoku Place, Hilo, HI 96720}
\altaffiltext{5}{Department of Physics and Astronomy, Johns Hopkins
  University, 3400 North Charles Street, Baltimore, MD 21218-2686}
\altaffiltext{6}{Steward Observatory, University of Arizona, 933 North
  Cherry Avenue, Tucson, AZ 85721}  
\altaffiltext{7}{Gemini Observatory, 670 North A'ohoku Place, Hilo, HI 96720}
\altaffiltext{8}{Department of Astronomy and Astrophysics, Pennsylvania
  State University, University Park, PA 16802}
\altaffiltext{9}{Apache Point Observatory, 2001 Apache Point Road,
  P.O. Box 59, Sunspot, NM 88349}



\begin{abstract}
We have resolved the newly discovered T~dwarf SDSS~J153417.05+161546.1
into a 0.11\arcsec\ binary using the Keck sodium laser guide star
adaptive optics system.
With an integrated-light near-IR spectral type of T3.5$\pm$0.5, this
binary provides a new benchmark for studying the distinctive $J$-band
brightening previously noted among early and mid-T dwarfs, using two
brown dwarfs with different spectral types but having a common
metallicity and age and very similar surface gravities.
We estimate spectral types of T1.5$\pm$0.5 and T5.5$\pm$0.5 for the two
components based on their near-IR colors, consistent with modeling the
integrated-light spectrum as the blend of two components.
The observed near-IR flux ratios of \sdssbin\ are unique compared to all
previously known substellar binaries: 
the component that is fainter at $H$ and $\Kp$ is brighter at $J$.
This inversion of the near-IR fluxes is a manifestation of the $J$-band
brightening within this individual binary system.
Therefore, \sdssbin\ demonstrates that the brightening can be intrinsic
to ultracool photospheres (\eg, arising from cloud disruption and/or
rapid increase in cloud sedimentation) and does not necessarily result
from physical variations among the observed ensemble of T~dwarfs (\eg, a
range in masses, ages and/or metallicities).
We suggest that the apparently large amplitude of the $J$-band
brightening may be due to a high incidence of unresolved binaries and
that the true amplitude of the brightening phenomenon could be more
modest.  This scenario would imply that truly single objects in these
spectral subclasses are relatively rare, in agreement with the small
effective temperature range inferred for the L/T transition.
\end{abstract}



\section{Introduction}

For many decades, color-magnitude diagrams (CMDs) have provided
fundamental insights into the physics of stellar evolution and stellar
atmospheres.  In an analogous fashion, the CMDs of brown dwarfs are
beginning to shed light on the physical properties of these low mass
objects, as an increasing number of distances have been determined from
trigonometric parallaxes for field brown dwarfs
\citep{2002AJ....124.1170D, 2003AJ....126..975T, 2003PASP..115.1207T,
  2004AJ....127.2948V} and from brown dwarf companions around stars with
known distances \citep[e.g.][]{1988Natur.336..656B, 1995Natur.378..463N,
  1998Sci...282.1309R, 2000ApJ...531L..57B, 2001AJ....121.3235K,
  2001astro.ph.12407L, 2002ApJ...567L.133P, 2003ApJ...584..453F,
  2004ApJ...617.1330M}.

However, interpretation of brown dwarf CMDs is considerably less
developed than for the case of stellar CMDs.  Brown dwarfs continually
cool as they age, unlike stars which occupy a fixed position on the CMD
during the main-sequence phase.  Therefore, there are degeneracies in
determining the masses and ages of brown dwarfs from the CMD.
Substellar photospheres have a rich appearance due to the presence of
dust and molecules (\eg, metal hydrides, water vapor, and methane), and
spectroscopic methods to determine the associated physical properties,
such as effective temperature and surface gravity, are still under
development \citep[e.g.][]{kirk05}.  The optical and infrared CMDs of
brown dwarfs show considerable scatter which could be due to differences
in metallicity, surface gravity (\ie, mass), rotation, dust properties,
and/or unrecognized binarity among the observational ensemble
\citep[e.g.][]{leg01, 2004AJ....127.3553K}.

Study of ultracool binary systems is a promising means to address these
issues, since binaries provide systems with common ages and
metallicities.  About two dozen L~dwarf binaries are now known from high
resolution imaging of over a hundred L~dwarfs, mostly by \HST\
\citep{1999ApJ...526L..25K, 1999Sci...283.1718M, 2001AJ....121..489R,
  2003AJ....125.3302G, 2003AJ....126.1526B, 2004AJ....128.1733G}.  In
contrast, only five ultracool binaries that include at least one T~dwarf
have been identified: 2MASS~J1225$-$2739AB (T6+T8;
\citealp{2003ApJ...586..512B}), 2MASS~J1534$-$2952AB (T5.5+T5.5;
\citealp{2003ApJ...586..512B}), $\epsilon$~Ind~Bab (T1+T6;
\citealp{2004A&A...413.1029M}), SDSS~J0423$-$0414AB (L6+T2;
\citealp{2005astro.ph.10580B}), and 2MASS~J2252$-$1730AB (L6+T2;
\citealp{2006ApJ...639.1114R}).\footnote{Some other objects are
  suspected of being L+T binaries, \ie, systems composed of an L~dwarf
  and a T~dwarf, based on their resolved optical data (2MASS~J0850+1057
  [\citealp{2001AJ....121..489R}] and 2MASS~J1728+3948
  [\citealp{2003AJ....125.3302G}]), near-IR flux ratios (Gl~337CD;
  \citealp{2005AJ....129.2849B}), or integrated-light spectra
  (2MASS~J0518$-$2828 [\citealp{2004ApJ...604L..61C}] and
  2MASS~J0920+3517 [\citealp{2001AJ....121..489R, 2001ApJ...561L.119N,
      2005astro.ph.10090B}]).}
Understanding the transition between objects of spectral type~L, marked
by very red IR colors and metal hydride absorption, and those of
spectral type~T, marked by blue IR colors and methane absorption, has
been an active area of inquiry \citep[e.g.][]{2004astro.ph..9389L}.  An
expanded sample of L-type and T-type binaries should shed light
on this issue.

We are conducting an imaging survey of nearby brown dwarfs using the new
Keck laser guide star adaptive optics system (LGS AO).  LGS AO enables
near diffraction-limited imaging over most of the sky with a 10-meter
telescope.  Thus, LGS AO is a powerful new capability for finding and
characterizing substellar binaries.  We have previously used Keck LGS AO
to discover that the nearby L~dwarf Kelu-1 is a binary system
\citep{2005astro.ph..8082L}, thereby explaining the unusual properties
of Kelu-1 (high luminosity, very red color, high inferred effective
temperature, and low lithium absorption line strength) compared to other
early L~dwarfs.  In this paper, we present the discovery of the binarity
of SDSS~J153417.05+161546.1 (hereinafter \sdssbin), a recently
discovered T~dwarf with an integrated-light near-IR spectral type of
T3.5$\pm$0.5 \citep{chiu05}, and consider the implications for
understanding the L/T transition.


\section{Observations}

We observed \sdssbin\ on 2005~May~1~UT using the sodium LGS AO system
\citep{2004SPIE.5490..321B, 2004SPIE.5490....1W} of the 10-meter Keck II
Telescope on Mauna Kea, Hawaii.  We used the facility IR camera NIRC2
and the $J$ (1.25~\micron), $H$ (1.64~\micron), and \Kp\ (2.12~\micron)
filters from the Mauna Kea Observatories (MKO) filter consortium
\citep{mkofilters1, mkofilters2}.  We used NIRC2's narrow camera, which
produces a 0.00994\arcsec~pixel\perone\ scale and a 10.2\arcsec\ field of
view.  Conditions were photometric with excellent seeing conditions
during the night, better than 0.6\arcsec\ in the optical as reported
by the neighboring Keck~I Telescope.  The total setup time for the
telescope to slew to \sdssbin\ and for the LGS AO system to be fully
operational was 18~minutes.  \sdssbin\ was observed over an airmass range
of 1.14--1.24.

The LGS brightness, as measured by the flux incident on the AO wavefront
sensor, was equivalent to a $V\approx10.3$~mag star.  The LGS provided
the wavefront reference source for AO correction, with the exception of
tip-tilt motion.  Tip-tilt aberrations and quasi-static changes in the
image of the LGS as seen by the wavefront sensor were sensed using the
$R=15.9$~mag star 1062-0236346 from the USNO-B1.0 catalog
\citep{2003AJ....125..984M}, located 14\arcsec\ away from \sdssbin.  The
overall image quality was somewhat worse compared to other data obtained
on the same night \citep{2005astro.ph..8082L} due to higher winds during
the \sdssbin\ observations.  The LGS AO-corrected images have full
widths at half maxima (FWHM) of 0.08\arcsec, 0.06\arcsec, and
0.07\arcsec\ at $JH\Kp$, respectively, with RMS variations of about
5--10\%.  The corresponding Strehl ratios were 0.02, 0.06, and 0.20
with RMS variations of about 5--20\%.
The FWHM and Strehl measurements were derived from the result of the
binary fitting process described below.

We obtained a series of five images in each filter, dithering the
telescope by a few arcseconds between each image.  The total on-source
integration time per filter was 7.5~minutes.  The sodium laser beam was
steered with each dither such that the LGS remained on \sdssbin\ for all
exposures.
This object was easily resolved into a binary system in all our data.
The images were reduced in a standard fashion.  We constructed flat
fields from the differences of images of the telescope dome interior
with and without lamp illumination.  Then we created a master sky frame
from the median average of the bias-subtracted, flat-fielded images and
subtracted it from the individual images. Images were registered and
stacked to form a final mosaic (Figure~1).  The known small geometric
distortions in NIRC2 have a negligible effect on our analysis.

To measure the flux ratios and relative positions of the two components,
we adopted two methods.  (1) The first method used an analytic model of
the point spread function (PSF) as the sum of two elliptical Gaussian
components, a narrow component for the PSF core and a broad component
for the PSF halo.  (2) The second method empirically determined the PSF
using the Starfinder software package \citep{2000A&AS..147..335D}, which
was designed for analysis of blended AO images.  For the individual
images obtained with each filter, we fitted for the flux ratio,
separation, and position angle of the binary.  The averages of the
results were adopted as the final measurements and the standard
deviations as the errors.  The results from the two different methods
were generally consistent.

In order to gauge the accuracy of these two measurement methods, we
created artificial binary stars from data of single stars that had
comparable image quality (similar tiptilt star properties, Strehl
ratios, and FWHM) to the \sdssbin\ data.  Our two methods were applied
to the artificial binaries to assess the fitting results and derived
errors.  Based on this analysis, we adopted the Starfinder measurements
for the $H$-band and \Kp-band data, and the Gaussian fitting results for
the $J$-band data.  These choices were not unexpected: the higher
Strehl, long wavelength data had a more structured PSF which was better
treated by Starfinder, while the much smoother, low Strehl $J$-band
images were modeled better by the two-gaussian PSF.  Table~1 presents
the resulting flux ratios and astrometry results.  

Table~2 reports the IR apparent magnitudes and colors of the individual
components, as derived from our flux ratio measurements and the
integrated IR magnitudes reported by \citet{chiu05}.  The absolute
magnitudes for \sdssbin\ are unknown, as a trigonometric parallax is not
available for this system.  Also, in principle variability could impact
the photometry determined for the individual components, since the
integrated-light photometry was obtained at a different epoch than the
Keck LGS AO measurements of the flux ratios.  IR monitoring of a limited
number of T~dwarfs does find low-level variability in some objects,
though the timescales and persistence are not well-constrained
\citep{2003AJ....126.1006E, 2004MNRAS.354..466K, 2005MNRAS.362..727K}.
\sdssbin\ has been observed at 4~epochs by the SDSS survey over a span
of 11~months and shows no $z$-band (0.9~\micron) variability at the
$\sim$0.1~mag level.

Our NIRC2 $J$ and $H$-band imaging used the same MKO filters as
\citet{chiu05} and hence computing magnitudes for the individual
components from our measured flux ratios is straight-forward.  However,
our NIRC2 data were obtained with the MKO \Kp-band (2.12~\micron) filter
in order to minimize the thermal background from the sky + AO system,
whereas the Chiu \etal\ photometry used the MKO $K$-band (2.20~\micron)
filter.  \citet{2005astro.ph..8082L} reported a polynomial fit for the
$(\Kp-K)_{MKO}$ color term as a function of spectral type for L~and
T~dwarfs using synthetic photometry based on the data from
\citet{2004AJ....127.3553K}.  However, for the range of spectral types
relevant to \sdssbin\ (see \S~3), the color term depends strongly on the
spectral type inferred from the $J-H$ color.  Instead, we use the
synthetic photometry to find a direct relation for the color term as a
function of the $J-H$ color:
  \begin{equation}
   (\Kp-K)_{MKO} = -0.100 + 0.143 \times (J-H)_{MKO},
  \end{equation}
valid for L~and T~dwarfs with $-0.5<(J-H)<1.0$.  The resulting $(\Kp-K)$
terms are 0.02 and $-$0.13~mags for components A and B,
respectively. The RMS scatter about the fit is 0.03~mags, which we add
in quadrature when determining the $K$-band photometry errors of the
individual components.

Figure~1 shows the unique behavior of the near-IR flux ratios for
\sdssbin{AB} compared to all other known substellar binaries: namely,
the identification of the brighter component depends on the observing
wavelength.  We designate the eastern object as the ``A'' component,
based on the fact that its inferred spectral type is earlier than the
western component, as discussed in \S~3.  The eastern object is brighter
in $H$ and $\Kp$ bands, and we would also expect it to be the brighter
component at optical wavelengths, based on the T~dwarfs with known
distances \citep{2003AJ....126..975T}.  The western, later-type object
is brighter only in the $J$-band images.  This can be understood as a
manifestation of the $J$-band brightening seen in the absolute
magnitudes of T~dwarfs, wherein objects of spectral types from about T2
to T5 are notably brighter at $J$-band compared to earlier type objects
\citep{2002AJ....124.1170D, 2003AJ....126..975T, 2004AJ....127.2948V}.
We discuss this further in \S~4.

No other sources are detected in our images.  For objects at
$\gtrsim$0.3\arcsec\ from \sdssbin, the final mosaics reached 
point source detection limits of 21.0, 21.0, and 20.7~mags at $JH\Kp$,
respectively, out to a separation of 3\arcsec.  For an age range of
0.5--5~Gyr and a estimated distance of 36~pc (see \S~3), these detection
limits correspond to about 10--30~\Mjup\ companions around
\sdssbin\ based on the models of \citet{2003A&A...402..701B}.


\section{Results: Properties of \sdssbin{AB}}

With only one epoch of imaging, we cannot ascertain whether the two
objects in our Keck images are physically associated based solely on
common proper motion.  However, the combination of the small proper
motion and the distinctive optical/IR colors can only be explained if
the system is a binary T~dwarf.  No relative motion was detected between
the two components during the 30~minutes elapsed time for the Keck LGS
AO imaging, so component~B cannot be a solar system body (which would be
expected to show a relative motion of $\gtrsim$0.5\arcsec).
\sdssbin\ was detected at $J$ and $H$-bands by 2MASS in 1999 and by SDSS
in both 2004 and 2005, giving a proper motion of 0.15\arcsec/yr at
position angle 248\degs\ east of north.  The small proper motion
indicates that the optical and IR photometry in \citet{chiu05}
represents the combined light of both objects, and thus our Keck IR flux
ratio measurements can be used to derive the resolved photometry of the
individual components.
The resulting IR colors for the two components are entirely consistent
with those of T~dwarfs.  Component B does have neutral $JHK$ colors
similar to hot O stars and white dwarfs, but if it were either of these
objects, the integrated optical light of the system would be
predominantly blue rather than very red as observed
($z-J=3.61\pm0.10$~mags; \citealp{chiu05}).
Hence, we conclude that the two objects constitute a binary T~dwarf.

To infer spectral types, we compare the resolved $JHK$ colors of
\sdssbin{AB} to those of ultracool dwarfs from
\citet{2004AJ....127.3553K} and \citet{chiu05}, excluding any known
binaries.  Our measurements and the published photometry were both
obtained with the MKO filter system.  For the L~dwarfs, we use the
published spectral types, which are based on the \citet{geb01} near-IR
classification scheme.  For the T~dwarfs, we use near-IR classfications
from the \citet{2005astro.ph.10090B} scheme.  (We adopt these two
classification schemes throughout this paper, which includes updating
older published T dwarf types to the new Burgasser \etal\ scheme.)  We
assume that the components of \sdssbin\ are themselves single, not
unresolved binaries.  Figure 2 shows that component~A has IR colors most
typical of the known T1 dwarfs, with T2 also being possible.  (The
colors of component~A are also similar to the few late-L dwarfs with
unusually blue near-IR colors in Figure~2.  However, as discussed below,
modeling of the integrated-light spectrum indicates that this is
unlikely.)  Component~B has IR colors most similar to the T5 dwarfs and
some of the T6~dwarfs.  We estimate spectral type estimates of
T1.5$\pm$0.5 and T5.5$\pm$0.5 for the two components.\footnote{We assume
  that the metallicity of \sdssbin\ is not very different than the bulk
  of the known T~dwarfs.  The low proper motion argues against it being
  a halo object.  Likewise, the one candidate low-metallicity T~dwarf
  reported by \citealp{2003ApJ...594..510B} is distinguished by an
  exceptionally blue $H-K$ color, whereas the colors of \sdssbin\ are
  consistent with other T~dwarfs.}

We modeled the integrated-light spectrum of \sdssbin\ as the blend of
two template T~dwarfs, in order to check the consistency with the flux
ratios and spectral types inferred from our Keck LGS imaging.  For
templates, we chose T~dwarfs from Figure~2 with complete $JHK$ spectra
and having similar near-IR colors to the two components of
\sdssbin\ (color differences of $\lesssim$0.1~mags).  Then we scaled the
template spectra to match the observed $J$-band flux ratio, summed the
templates to form a composite spectrum, and scaled the resulting
composite to the observed integrated-light spectrum.  (Note that this
approach has only one free parameter in the modeling, namely the
distance to the binary.)  The comparisons between the model composite
spectra and the observed spectrum show good agreement
(Figure~3).\footnote{We also tried using T~dwarfs with known distances
  as templates (as compiled in \citealp{2004AJ....127.3553K}), directly
  summing the flux-calibrated spectra without imposing the $J$-band flux
  ratio.
The resulting set of blended spectra could not simultaneously match the
\sdssbin\ integrated-light spectrum and the observed $JHK$ flux ratios.
The mismatch is not surprising given that most of the T~dwarfs with
measured distances do not have the same near-IR colors as
\sdssbin{A}~and~B.  As Figure~2 shows, there is significant color
scatter within a given T~subclass.}  Likewise, although \sdssbin{A} has
near-IR colors similar to some of the bluest late-L dwarfs, it was not
possible to create a well-matched composite spectrum with a blue late-L
dwarf as a template, and thus we discount this possibility.  Overall,
the spectral types estimated from modeling the integrated-light spectrum
agree well with those from the resolved Keck near-IR colors.

In the absence of a trigonometric parallax, the model composite spectra
can provide a rough distance estimate for \sdssbin{AB}.  The spectra
were created using 2MASS~J2356$-$1553 (T5.5; \citealp{burg01}) and
SDSS~J2124+0100 (T5; \citealp{chiu05}) to represent component~B.
\citet{2004AJ....127.2948V} have measured a trigonometric parallax for
2MASS~J2356$-$1553.  For SDSS~J2124+0100, we assumed an absolute
$J$-band magnitude of 14.37~mags, based on the polynomial fit for the
absolute $J$-band magnitude as a function of spectral type from
\citet{2004AJ....127.3553K}.  Assuming these two mid-T templates are
single objects, the inferred distance modulus for \sdssbin{AB} is
2.7--2.9~mags ($\approx$36~pc).
To check on the distance estimated from our spectral modeling, we
applied the same procedure to the integrated-light spectrum of the
binary T~dwarf \eInd~{Bab} \citep{2004A&A...413.1029M} and derived a
distance estimate of 3.3~pc, in good agreement with the measured
distance of 3.626$\pm$0.009~pc \citep{1997A&A...323L..49P}.


We can estimate the mass ratio \sdssbin{AB}, independent of its
distance.  For the assumed spectral types of T1.5$\pm$0.5 and
T5.5$\pm$0.5, the difference in bolometric corrections for the two
components is 0.55$\pm$0.15~mags, with $BC_K$ being larger for the
T1.5$\pm$0.5 component \citep{gol04}.  The $K$-band flux ratio for the
system is 1.21$\pm$0.05~mags, based on our Keck LGS images and the
($\Kp-K$) color term calculated in \S~2.  Hence, the difference in
bolometric luminosities is 0.7$\pm$0.2~mags, with component~A having the
higher luminosity.  \citet{bur01} provide an approximate mass-luminosity
relation for coeval solar-metallicity brown dwarfs, $L_{bol} \propto
M^{2.64}$.  This implies a mass rataio of about 0.80$\pm$0.05 for the
two components of \sdssbin, which is similar to the nearly equal mass
ratios found for other binary brown dwarfs
\citep[e.g.][]{2003ApJ...586..512B}.

Likewise, we can estimate the surface gravity difference for the two
components, independent of the distance.  Since the radii of brown
dwarfs older than $\gtrsim$300~Myr are nearly independent of the masses,
the near-unity mass ratio means that the surface gravities of
\sdssbin{A} and~B are very similar.  \citet{bur01} present analytic fits
to substellar evolutionary models that show the quantity
$g^{1.7}\Teff^{-2.8}$ is constant at fixed age.  Adopting \Teff\ ranges
of $\approx$1300--1400~K and $\approx$1100--1200~K for components~A
and~B, respectively (\citealp{gol04}, which assumes an age of 3~Gyr),
the inferred difference in surface gravities is approximately
0.05--0.15~dex.


We can make a rough estimate of the orbital period of \sdssbin{AB},
assuming that the true semi-major axis is not very different than the
projected separation.  (For example, \citealp{1992ApJ...396..178F} show
for binaries with a random distribution of orbital elements that on
average the true semi-major axis is about 1.3$\times$ larger than the
projected semi-major axis.)
To estimate the total system mass, we use the 36~pc distance inferred
from the composite spectra fitting and the aforementioned $BC_K$ to
estimate bolometric luminosities of $2.2\times10^{-5}$ and
$1.2\times10^{-5}$~\Lsun\ for the two components.  Models by
\citet{bur01} give total masses for the system of about 60~\Mjup,
85~\Mjup, and 145~\Mjup\ for assumed ages of 0.5, 1.0, and 5.0~Gyr.
For an assumed semi-major axis of 4.0~AU, these mass estimates lead to
an expected orbital period of 22--33~years.  More generally,
\citet{1999PASP..111..169T} shows that $\approx$85\% of randomly
oriented orbits have a true semi-major axis of 0.5--2.0$\times$ the
projected separation, corresponding to a range of orbital periods of
$\approx$8--95~yr for \sdssbin{AB}.


\section{Discussion}

\subsection{The Nature of the L/T Transition}

The rapid changes in near-IR colors at the L/T transition are generally
understood to be due to a change in the photospheric dust content, but
the very abrupt nature of this effect over a small inferred range in
effective temperature remains a challenge to theoretical models.  One
key observational characteristic of the L/T transition is the unusual
strong ($\sim$1~mag) brightening of the $J$-band flux among the early
and mid-T~dwarfs compared to the late-L~dwarfs
\citep{2002AJ....124.1170D, 2003AJ....126..975T, 2004AJ....127.2948V}.
This brightening is also seen at far-red ($I$ and $Z$) and other near-IR
wavelengths ($H$ and $K$), though the effect is greatest at $J$-band.
This wavelength dependence is expected, since at $J$-band the
photospheric opacities are most sensitive to the properties of the
clouds \citep{2005astro.ph..9066B}.

Models can qualitatively accomodate the observed transition from the
very red L~dwarfs to the very blue T~dwarfs \citep{2002ApJ...568..335M,
  2002ApJ...575..264T, 2005astro.ph..9066B}.  But no model thus far has
succcessfully reproduced the observed $J$-band brightening along a
single evolutionary track.  Based on a suggestion by
\citet{2001ApJ...556..872A}, \citet{2002ApJ...571L.151B} show that
disruption of the condensate clouds at fixed \Teff\ could produce an
enhancement of the $J$-band flux, as holes in the cloud deck reveal the
deeper, hotter regions of the photosphere.  \citet{2004AJ....127.3553K}
speculate that a rapid increase in the dust sedimentation efficiency
over a small range of \Teff\ could be responsible.  However, as pointed
out by \citet{2005astro.ph..9066B}, these explanations are ad hoc,
lacking a firm physical basis for the rapid transition.

A very different picture is suggested by \citet{2003ApJ...585L.151T}.
They propose that the $J$-band brightening does not occur along the
evolutionary track of a given object.  Rather, in their ``unified cloudy
model'' the \Teff\ at which the L~dwarf sequence merges with the T~dwarf
sequence depends on surface gravity: for younger or lower-mass objects,
the L/T transition occurs at younger ages and brighter $J$-band absolute
magnitudes than for older or higher-mass objects.  Thus, in their models
the $J$-band brightening arises from the mass and age spreads within the
known ensemble of L~and T~dwarfs and does not occur during the evolution
of a single object.  Objects that comprise the $J$-band brightening are
predicted to be younger and to have lower surface gravities than objects
which do not \citep[c.f.][]{2003AJ....126..975T}.

The flux crossover in the near-IR that we observe in \sdssbin{AB}
provides a benchmark for understanding the $J$-band brightening, as the
binary presumably constitutes a system of common age and metallicity and
nearly identical surface gravity.  This crossover cannot be explained by
the aforementioned unified cloudy model of \citet{2003ApJ...585L.151T}.
Their model isochrones show that no $J$-band brightening occurs for a
coeval population, in disagreement with our observations of
\sdssbin{AB}.  The fact that \sdssbin{AB} displays the $J$-band
brightening between its two coeval components reveals that the
phenomenon can originate from processes intrinsic to ultracool
photospheres with very similar surface gravities, such as the
aforementioned cloud disruption and/or rapid dust settling
scenarios.\footnote{The reappearance of KI and FeH absorption in the
  far-red spectra of T2--T5 dwarfs, after having weakened among the
  late-L dwarfs \citep{2002ApJ...571L.151B, 2004ApJ...607..499N,
    2005ApJ...623.1115C}, also argues that true photospheric processes
  are involved in the $J$-band brightening.}  Similarly, models by
\citet{2005astro.ph..9066B} indicate that variations in metallicities
among the observed~L and T~dwarfs could be responsible for the $J$-band
brightening (though this is not the preferred solution of Burrows
\etal).  \sdssbin{AB} demonstrates that objects of the same metallicity
can exhibit the $J$-band brightening.

A good point of comparison is the T~dwarf binary \eInd~Bab
\citep{2004A&A...413.1029M}.  \citet{2003A&A...398L..29S} determined an
integrated-light spectral type of approximately T2.5 using the
\citet{geb01} classification scheme; we analyzed the published spectrum
and found that the type is unchanged with the newer
\citet{2005astro.ph.10090B} scheme.
The individual components of this 0.7\arcsec\ (2.6~AU) binary have
spectral types of~T1 and~T6 on the Burgasser \etal\ scheme.  The near-IR
colors are consistent with \eInd~Ba having a comparable spectral type to
\sdssbin{A} and \eInd~Bb being about one~subclass later than \sdssbin{B}
(Figure~2).\footnote{We recomputed photometry for the individual
  components of \eInd~Bab.  The \citet{2004A&A...413.1029M}
  determination assumed that the VLT/NACO and 2MASS filter systems were
  identical.  Based on synthetic photometry, we determined the (small)
  color terms between the two filter systems specific for T~dwarfs and
  then used the published measurements to re-compute the colors:
  $(J-H)_{MKO} =\{0.53,-0.34\}$, $(H-K)_{MKO} = \{0.18, -0.34\}$, and
  $(J-K)_{MKO} = \{0.71,-0.69\}$ for components~A and~B, respectively.
  The resulting colors are $\le$0.03~mags different from the published
  values.}  However, the flux ratios of \eInd~Bab do not show the
crossover seen between the T1.5 ad T5.5 components of \sdssbin{AB}.
This suggests that by spectral type T6 the $J$-band brightening no
longer occurs, consistent with the observed CMD of T~dwarfs (and
ignoring possible complications due to surface gravity effects, as the
\eInd\ system is somewhat younger [1--2~Gyr] than expected for field
objects).


\subsection{Is the $J$-Band Brightening Enhanced by Unrecognized Binarity?}

The flux crossover seen in the near-IR for \sdssbin{AB} demonstrates
that the $J$-band brightening is a true physical effect and not solely a
manifestation of large physical variations among the L~and T~dwarfs with
known distances (unless \sdssbin{B} itself is an unresolved binary).
However, the integrated-light photometry of \sdssbin\ is in accord with
the other $J$-band brightening objects (Figures 2, 4, and 5), most of
which have not yet been imaged at high angular
resolution.\footnote{Likewise, the integrated-light spectrum of
  \sdssbin\ shows no strong differences from the spectra of the other
  two known T3.5 objects, SDSS~J1750+1759 \citep{geb01} and
  SDSS~J1214+6316 \citep{chiu05}.}  This suggests that unrecognized
binarity among early/mid-T~dwarfs may strongly affect the appearance of
the $J$-band brightening in the brown dwarf CMD.
Furthermore, \citet{2005astro.ph.10580B} have recently noted that
objects with spectral types of L7--T2 appear to have a higher binary
frequency relative to earlier-type L~dwarfs and mid/late-type T~dwarfs
(though a complete analysis of the selection biases is still needed).
This finding covers an earlier spectral type range than the $J$-band
brightening objects, but it does provide more evidence that binarity
likely plays an important role in the L/T transition.

To examine this notion further, we consider here a scenario in which
truly single $J$-band brightening objects are in fact rare.
In this situation, it is the blended light of two components with
different spectral types that produces an integrated-light spectral type
of $\approx$T2--T4.5.  Thus, the binary frequency of these spectral
subclasses is greater compared to other T~dwarfs, and the apparent
amplitude of the $J$-band brightening effect is enhanced.
\citet{2004AJ....127.2948V} suggested that the $J$-band brightening
cannot be due to unresolved binarity, because a significant number of
objects contribute to the effect.  However, their implicit assumption is
that a given spectral subclass contains a mix of singles and binaries.
Their objection would not be relevant in our scenario, wherein the
reason that most objects are classified as T2--T4.5 is because they are
binaries.

The observed CMD of T~dwarfs lends some support to the idea that
binarity might enhance the amplitude of the $J$-band brightening.
Unrecognized binarity can add up to 0.75~mag of scatter in the absolute
magnitudes, in addition to scatter caused by other effects, \eg,
variations in surface gravity \citep{2004AJ....127.3553K,
  2005astro.ph..9066B}.  Therefore, we would naturally expect at least
this much scatter in the absolute magnitudes at a given spectral
subclass, and this is largely observed (\eg, Figure 4 in this paper;
Figure~2 of \citealp{2004AJ....127.2948V}; Figure~9 of
\citealp{2003AJ....126..975T}; Figure~8 of
\citealp{2004AJ....127.3553K}).  However, the $J$-band brightening
objects appear to have a {\em smaller} scatter in their absolute
magnitudes compared to earlier and later spectral subclasses.  This
reduced scatter could be explained if in fact most $J$-band brightening
objects are binaries, with truly single objects being rare.  (An
alternative explanation would be that the reduced scatter arises from a
selection effect in the targets chosen by the parallax programs.
However, it is not obvious what effect would act only on the $J$-band
brightening objects and not the other subclasses.)

A similar hint, in a slightly different form, is seen in the effective
temperatures of L~and T~dwarfs derived by \citet{gol04}.  Their results
suggest that T2--T4.5 dwarfs may have slightly higher temperatures than
the late-L and T0--T1 dwarfs (see their Figure~6).  While this
discrepancy is within the uncertainties in the age estimates (which
impact the model-derived radii used in the \Teff\ calculation), the
systematically higher \Teff's are a curious trend in the data,
especially under the assumption that the spectral type scale is a proxy
for the \Teff\ scale \citep[c.f.][]{2005ApJ...621.1033T}.
One possibility would be that the T2--T4.5 dwarfs in Golimowski
\etal\ have systematically younger ages than the earlier or later-type
objects, but again it is not obvious how such a selection bias would
occur.  A more natural explanation is that the elevated \Teff's result
from unrecognized binarity.
This is amply demonstrated by the L3~dwarf Kelu-1 and the T0~dwarf
SDSS~J0423$-$0414 --- their effective temperatures appear to to be
anomalously large compared to objects of similar spectral types, but
these deviations are in fact an artifact of previously unrecognized
binarity \citep{2005astro.ph..8082L, 2005astro.ph.10580B}.  For the case
of equal-luminosity components, \Teff\ will be overestimated by about
20\% ($2^{1/4}$); applying this correction to the Golimowski
\etal\ T2--T4.5 dwarfs would bring their temperatures into agreement with
the earlier-type objects.\footnote{By the same rationale,
  SDSS~J0032+1410 (L8) and 2MASS~J0328+2302 (L9.5) are candidates for
  being unresolved binaries based on their relatively high \Teff's.}

In fact, if all the L~and T~dwarfs with anomolously high temperatures in
the Golimowski \etal\ sample are actually unresolved binaries, we would
infer that the $J$-band absolute magnitudes across the L/T transition
are relatively constant (Figure~4).  Similarly, the inflections in the
brown dwarf CMDs at other wavelengths would also be impacted.
Accounting for the possible binaries, Figure~5 shows that the apparent
plateau in the $H$ and $K$-band fluxes of the early/mid-T~dwarfs may be
lessened, leading to a more monotonic dependence on on the spectral
type.  Table~3 provides polynomial fits for the absolute magnitude as a
function of spectral type, with and without the possible binaries.

At face value, a higher binary frequency among the $\approx$T2--T4.5
objects relative to earlier and later-type objects may seem physically
implausible.  However, the binary frequency as a function of spectral
type is not the same as the frequency as a function of \Teff.  While the
frequency as a function of \Teff\ is expected to change smoothly, the
frequency as a function of spectral type may not show the same behavior,
since it also depends on how the light of individual binary components
blends together.
In particular, the large changes in spectral morphology and the
non-monotonic behavior of the absolute magnitudes during the L/T
transition may lead to interesting observational consequences when these
objects form binaries.\footnote{Since the IR absolute magnitudes around
  the L/T transition do not decrease monotonically as a function of
  spectral type, unresolved binaries in the L/T transition region with a
  range of mass ratios can be significantly displaced in the CMD in both
  absolute magnitude and integrated-light spectral type compared to the
  individual components.  In contrast, if the magnitude-spectral type
  relation were monotonic and very steep, then a population of
  unresolved binaries would mostly lead to a spread in the magnitudes,
  due to the equal-mass binaries; the unequal-mass binaries would not
  have much impact on the CMD, since the secondary components would not
  substantially contribute to the integrated-light properties.}  For the
sake of illustration, if we consider an extreme example where truly
single T2--T4.5 objects do not exist at all, the \sdssbin{AB} (T3.5) and
\eInd~Bab (T2.5) systems demonstrate that these subclasses can be
populated with binaries composed of an early and a late-type T dwarf --
in this case, the T2--T4.5 dwarfs would have a 100\% binary fraction,
even if the earlier and later-type T~dwarfs have much lower binary
fractions.

A relative paucity of single objects in the $J$-band brightening regime
can be naturally explained if the corresponding \Teff\ range is very
small, \eg, as suggested by \citet{2000AJ....120..447K} and indicated by
measurements from \citet{2004AJ....127.2948V} and \citet{gol04}.  The
small \Teff\ range would correspond to a relatively short-lived
evolutionary phase, and therefore relatively few substellar objects
would be found in the corresponding spectral subclasses.
However, even if truly single $\approx$T2--T4.5 objects are rare, a
higher binary frequency among these objects may not be apparent in the
spectral type distribution of objects found by 2MASS and SDSS.  In such
magnitude-limited surveys, the effective detection volume for binary
stars is larger than for single stars, and thus a survey's observed
spectral type distribution depends on several factors --- the true
binary frequency, the slope of the relationship between absolute
magnitude and spectral type, and the effect of unresolved binarity on
spectral types determined from integrated-light photometry and/or
spectroscopy.  A Monte Carlo simulation, beyond the scope of this paper,
would be valuable in understanding the interplay of these effects.

Given that only six~T2--T4.5 dwarfs have measured distances, the
discussion here is inevitably speculative --- such a small sample
hinders more detailed considerations.\footnote{The six T2--T4.5~dwarfs
  with known distances are SDSS~J1254$-$0122 (T2), \eInd~{Bab} (T2.5 in
  integrated light), SDSS~J1021$-$0304 (T3), SDSS~J1750+1759 (T3.5),
  2MASS~J0559$-$1404 (T4.5), and SDSS~J0207+0000 (T4.5) --- see
  compilation in \citet{2005astro.ph.10090B}.  Also, SDSS~J0423$-$0414
  (T0 in near-IR integrated light) has recently been resolved into a
  binary by \citet{2005astro.ph.10580B}; the components have estimated
  spectral types of L6$\pm$1 and T2$\pm$1 based on their near-IR fluxes
  and modeling of the combined-light spectrum.  Resolved $JHK$
  magnitudes are not available for this system, so we estimate them for
  Figures~4 and~5 by scaling template spectra of the individual
  components to the resolved optical fluxes.}  Similarly, our adopted
spectral range of $\approx$T2--T4.5 for the strong $J$-band brightening
phenomenon should be considered as approximate.  The T~dwarf binaries
\sdssbin{AB}, \eInd~Bab, and SDSS~J0423$-$0414AB all have components
with estimated spectral types of T1--T2.  Unless these components are
themselves binary, this shows that single early-T~dwarfs are not
uncommon.  Indeed the positions of these components in brown dwarf CMDs
are quite similar (Figures~4 and~5) and are somewhat fainter than the
T2~dwarf SDSS~J1254$-$0122, which thus far appears to be a single object
\citep{2003ApJ...586..512B}.

If binarity does play a significant role in the $J$-band brightening
objects, there are some natural consequences.  (1) More extensive
imaging surveys with \HST\ and LGS AO should find a greater binary
frequency among these objects compared to earlier and later-type
objects, after accounting for selection effects inherent in current
magnitude-limited samples of brown dwarfs.  Of course, some systems may
have projected separations below the resolution limit of imaging surveys
and would only be detectable as spectroscopic binaries
\citep[e.g.][]{2005MNRAS.362L..45M}.
(2) Such binaries should often be composed of an earlier-type object,
say late-L to early-T, and an object of approximately T5 or later.
(3) More generally, the $J$-band brightening objects should have
relatively lower space densities compared to other ultracool dwarfs.
In this regard, a populous volume-limited sample of L~and T~dwarfs with
trigonometric parallaxes would be valuable for studying the nature of the
L/T transition.  Such a sample is unlikely to be achieved with current
parallax programs, which can only target a limited number of
pre-selected objects.  The upcoming Pan-STARRS project
\citep{2002SPIE.4836..154K} will monitor the entire sky observable from
Hawai`i with high astrometric precision (planned 1-d RMS astrometric
error of 10~mas per observing epoch).  The resulting dataset should be a
promising means to assemble a large sample of brown dwarfs with accurate
distances.


\section{Summary}

We have obtained high angular resolution imaging of the T3.5~dwarf
\sdssbin\ using the Keck sodium LGS AO system.  \sdssbin\ is resolved
into a 0.11\arcsec\ binary, and its low proper motion and distinctive
optical/IR colors indicate that it is a physical pair.  The resolved
near-IR colors give estimated spectral types of T1.5$\pm$0.5 and
T5.5$\pm$0.5, consistent with modeling the integrated-light spectrum as
the blend of two components.  This analysis also produces an estimated
distance of 36~pc, based on the distances to the mid-T~dwarfs used in
the modeling.  The inferred (distance-independent) differences in the
masses and surface gravities of the two components are about 20\% and
0.1~dex, respectively.

SDSS~J1534+1615{AB} serves as a new benchmark for understanding the
$J$-band brightening seen among the early/mid-T~dwarfs.  The binary's
near-IR flux ratios show an inversion, wherein component~A is brighter
at $H$ and $K$-bands but component~B is brighter at $J$-band by about
0.2~mags.  Thus, this binary system embodies the $J$-band brightening
phenomenon seen in the CMD of T~dwarfs.  Since the components of
\sdssbin\ presumably have a common age, a common metallicity, and a very
similar surface gravity, our results demonstrate that the $J$-band
brightening is intrinsic to ultracool photospheres and not necessarily a
by-product of observing an ensemble of brown dwarfs with different
masses, ages, and/or metallicities.  Likewise, the brightening mechanism
must be able to operate over a relatively small range
($\approx$200~K) in effective temperature.  The magnitudes and
colors of \sdssbin{AB} and the binary T~dwarf \eInd~Bab are quite
similar.  However, \eInd~Bab, with resolved spectral types of T1 and T6,
does not show a crossover in its near-IR flux ratios, suggesting that
the $J$-band brightening may end by around spectral type~T5.

Based on the limited observations to date, we have considered the
possibility that the large amplitude of the $J$-band brightening among
early/mid-T~dwarfs is caused by a high frequency of unresolved binaries.
\sdssbin{AB} and \eInd~{Bab} provide examples of objects which appear to
reside at/near the CMD location of the $J$-band brightening, but once
their resolved properties are taken into account, the individual
components are consistent with early-T's and mid/late-T's.  We speculate
that the possibly smaller scatter seen in the infrared CMDs and the
slightly higher \Teff's inferred for the $\approx$T2--T4.5 dwarfs could
be hints of unresolved binarity, leading to an enhancement in the
apparent amplitude of the $J$-band brightening.  In this interpretation,
truly single $\approx$T2--T4.5 objects would be relatively rare, with
most objects of these spectral subclasses being composed of the blended
light of an earlier and later-type component.  This hypothesis is in
accord with the small \Teff\ range inferred for the late-L to mid-T
subclasses; photospheric dust removal seems to occur very rapidly during
the L/T transition.

Further study of \sdssbin{AB} will shed light on the nature of the L/T
transition and the $J$-band brightening.  A trigonometric parallax will
accurately determine the CMD positions of the combined system and
individual components.  Resolved spectroscopy of the two components will
allow for more accurate spectral classification.  Also, given the common
metallicity and similar surface gravities of the two components, such
data can isolate the temperature-sensitive and dust-sensitive features
in the spectra of the early/mid-T~dwarfs.  Given the estimated
$\sim$28~yr period, follow-up high angular resolution imaging over the
next several years can trace a significant fraction of the orbital
motion and lead to a dynamical mass estimate.  Ultimately, a much larger
sample of T~dwarf binaries from \HST, LGS AO, and IR radial velocities
will provide the means to fully explore the physical processes
encapsulated in the substellar color-magnitude diagram.


\acknowledgments

We gratefully acknowledge the Keck LGS AO team for their impressive
efforts in bringing the LGS AO system to fruition.  It is a pleasure to
thank Antonin Bouchez, David LeMignant, Marcos van Dam, Randy Campbell,
Robert LaFon, Gary Punawai, Peter Wizinowich, and the Keck Observatory
staff for assistance with the observations; John Rayner, Robert Jedicke,
Adam Burrows, Adam Burgasser, Neill Reid, and Marcos van Dam for
productive discussions; and Dagny Looper for a careful reading of the
manuscript.  We thank Alan Stockton for a fortuitous swap of observing
nights.  Our research has benefitted from the 2MASS data products;
NASA's Astrophysical Data System; the SIMBAD database operated at CDS,
Strasbourg, France; and the M, L, and T dwarf compendium housed at
DwarfArchives.org and maintained by Chris Gelino, Davy Kirkpatrick, and
Adam Burgasser \citep{2003IAUS..211..189K, 2004AAS...205.1113G}.  ML
acknowledges support for this work from NSF grant AST-0507833 and a
Sloan Research Fellowship.  XF is supported by NSF grant AST-0307384, a
Sloan Research Fellowship, and a Packard Fellowship for Science and
Engineering.
Funding for the SDSS and SDSS-II has been provided by the Alfred
P. Sloan Foundation, the Participating Institutions, the National
Science Foundation, the U.S. Department of Energy, the National
Aeronautics and Space Administration, the Japanese Monbukagakusho, the
Max Planck Society, and the Higher Education Funding Council for
England. The SDSS Web Site is http://www.sdss.org/.
The SDSS is managed by the Astrophysical Research Consortium for the
Participating Institutions. The Participating Institutions are the
American Museum of Natural History, Astrophysical Institute Potsdam,
University of Basel, Cambridge University, Case Western Reserve
University, University of Chicago, Drexel University, Fermilab, the
Institute for Advanced Study, the Japan Participation Group, Johns
Hopkins University, the Joint Institute for Nuclear Astrophysics, the
Kavli Institute for Particle Astrophysics and Cosmology, the Korean
Scientist Group, the Chinese Academy of Sciences (LAMOST), Los Alamos
National Laboratory, the Max-Planck-Institute for Astronomy (MPIA), the
Max-Planck-Institute for Astrophysics (MPA), New Mexico State
University, Ohio State University, University of Pittsburgh, University
of Portsmouth, Princeton University, the United States Naval
Observatory, and the University of Washington.
%
%
Finally, the authors wish to recognize and acknowledge the very
significant cultural role and reverence that the summit of Mauna Kea has
always had within the indigenous Hawaiian community. We are most
fortunate to have the opportunity to conduct observations from this
mountain.

\clearpage


\begin{figure}
\vskip -8in
\vskip 7.5in
\hskip 2.1in
\centerline{\includegraphics[width=6in,angle=90]{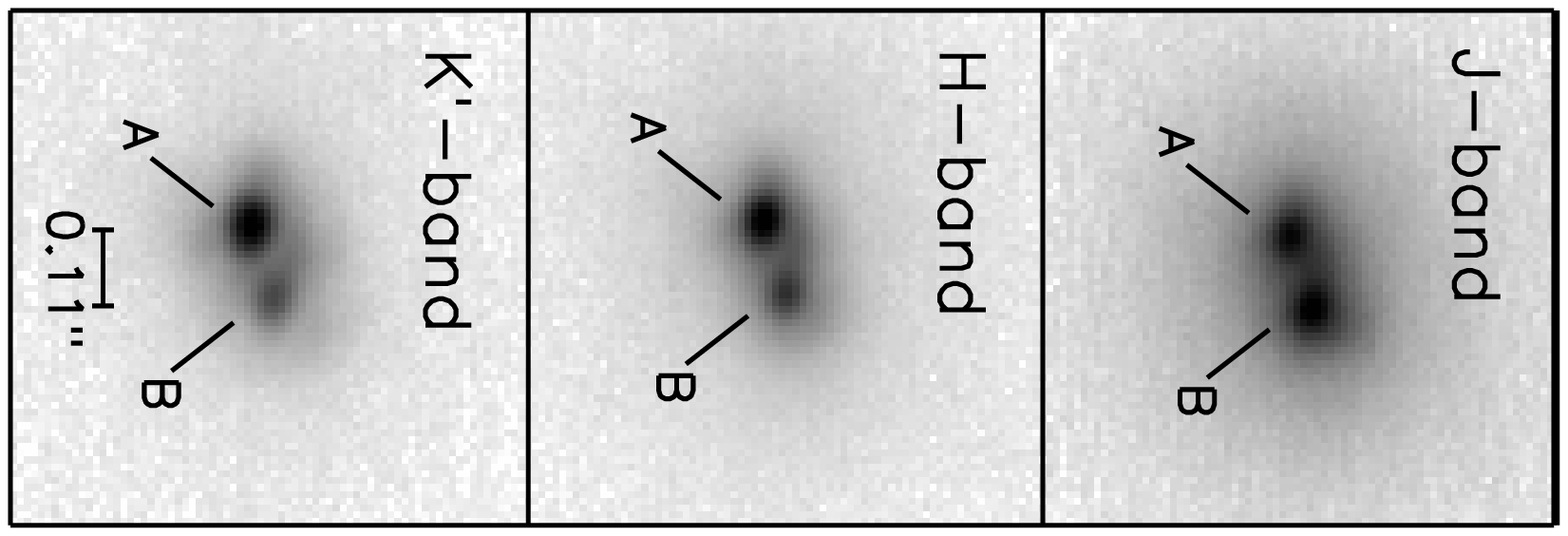}}
\vskip -6.02in
\hskip 4.2in
\centerline{\includegraphics[width=6in,angle=90]{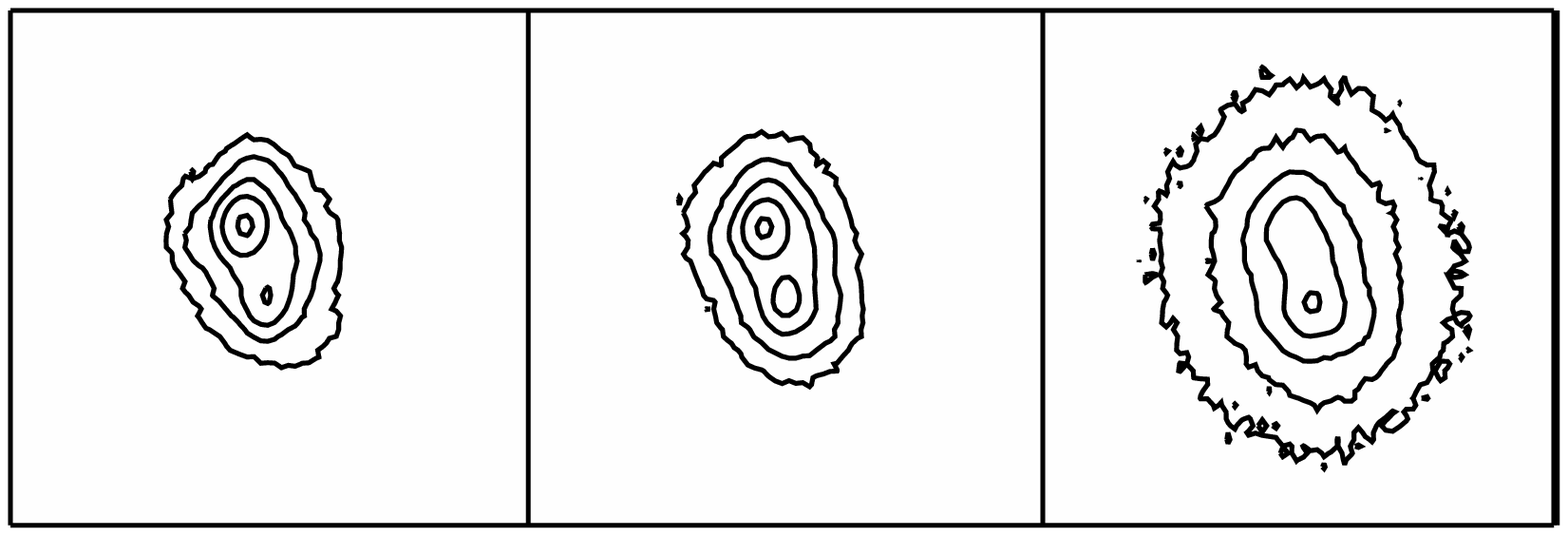}}
\vskip 2ex
\caption{\normalsize $JH\Kp$-band imaging of \sdssbin{AB} from Keck LGS
  AO.  North is up and east is left.  Each image is 0.75\arcsec\ on a
  side. The binary separation is 0.110\arcsec\ $\pm$ 0.005\arcsec.
  Contours are drawn from 90\%, 45\%, 22.5\%, 11.2\%, 5.6\% and 2.8\% of
  the peak value in each bandpass.  The source to the southeast (left)
  is component~A (brighter at $H$ and \Kp, fainter at $J$, spectral type
  T1.5$\pm$0.5), and the source to the northwest (right) is component~B
  (fainter at $H$ and \Kp, brighter at $J$, spectral type T5.5$\pm$0.5).
  The images are slightly east-west elongated due to telescope
  windshake.}
\end{figure}

\begin{figure}
\vskip -1.7in
\hskip 0.2in
\centerline{\includegraphics[width=6in,angle=90]{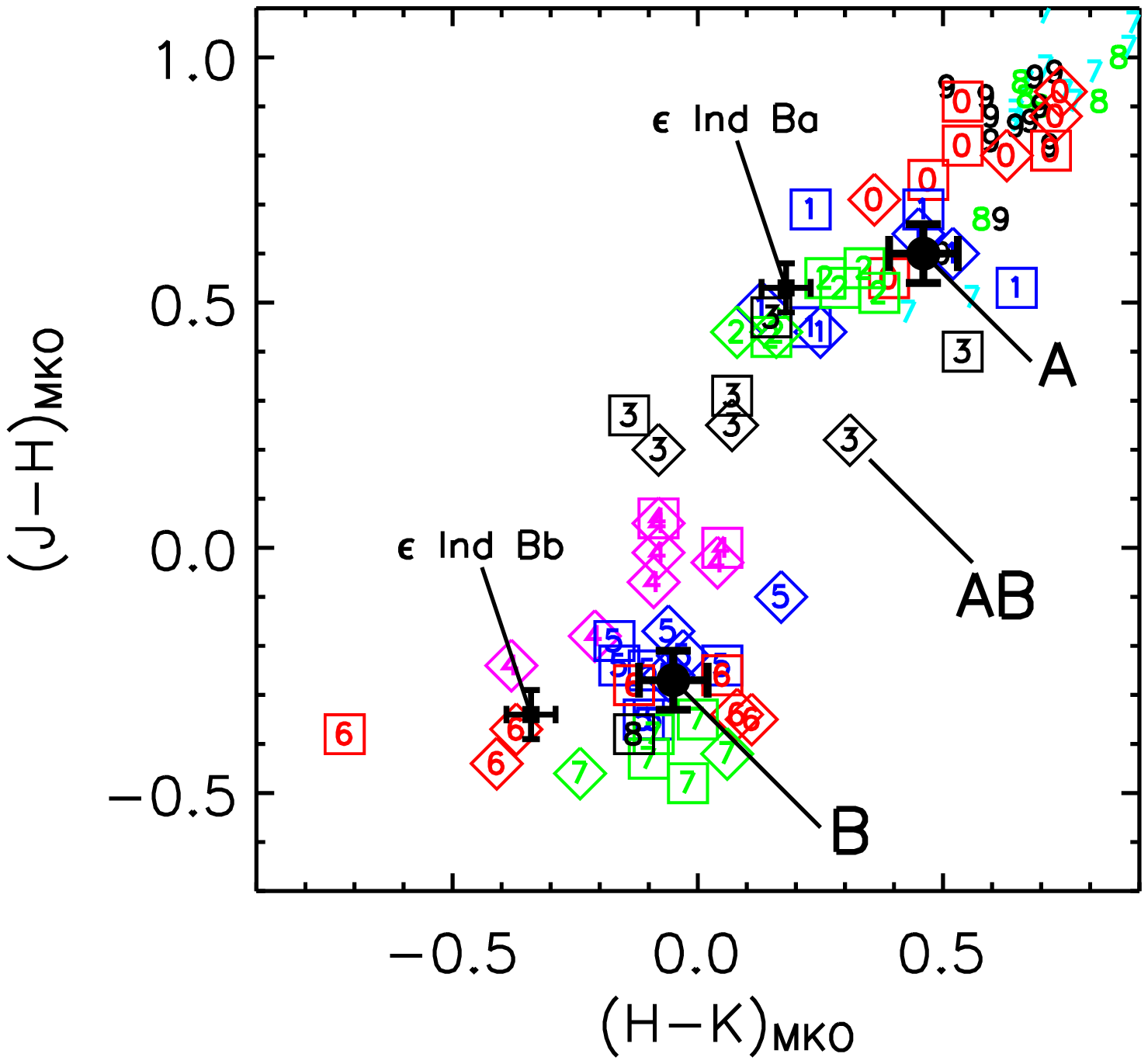}}
\vskip -2ex
\caption{\normalsize Near-IR colors of \sdssbin{AB} compared with nearby
  late-L~and T~dwarfs from \citet{2004AJ....127.3553K} and
  \citet{chiu05}.  The numbers indicate the near-IR spectral subclass of
  the objects, with half subclasses being rounded down (\eg, T3.5 is
  labeled as ``3'').  The late-L~dwarfs (classified on the
  \citealp{geb01} scheme) are plotted as bare numbers.  The T~dwarfs (on
  the \citealp{2005astro.ph.10090B} scheme) are plotted as circumscribed
  numbers, with squares for integer subclasses (\eg, T3) and diamonds
  for half subclasses (\eg, T3.5).  The photometry errors are comparable
  to or smaller than the size of the plotting symbols.  Known binaries
  are excluded except for \sdssbin{AB}, which is labeled on the plot as
  A, B, and AB.  The inferred spectral types for components~A and~B of
  \sdssbin\ are T1.5$\pm$0.5 and T5.5$\pm$0.5, respectively.  The
  individual components of \eInd~Bab are shown for comparison, where we
  have converted the published photometry to the MKO system (see \S~4.1)
  and assumed errors of 0.05~mags.
 \label{plot-sptype}}
\end{figure}

\begin{figure}
\centerline{\includegraphics[width=4.5in,angle=0]{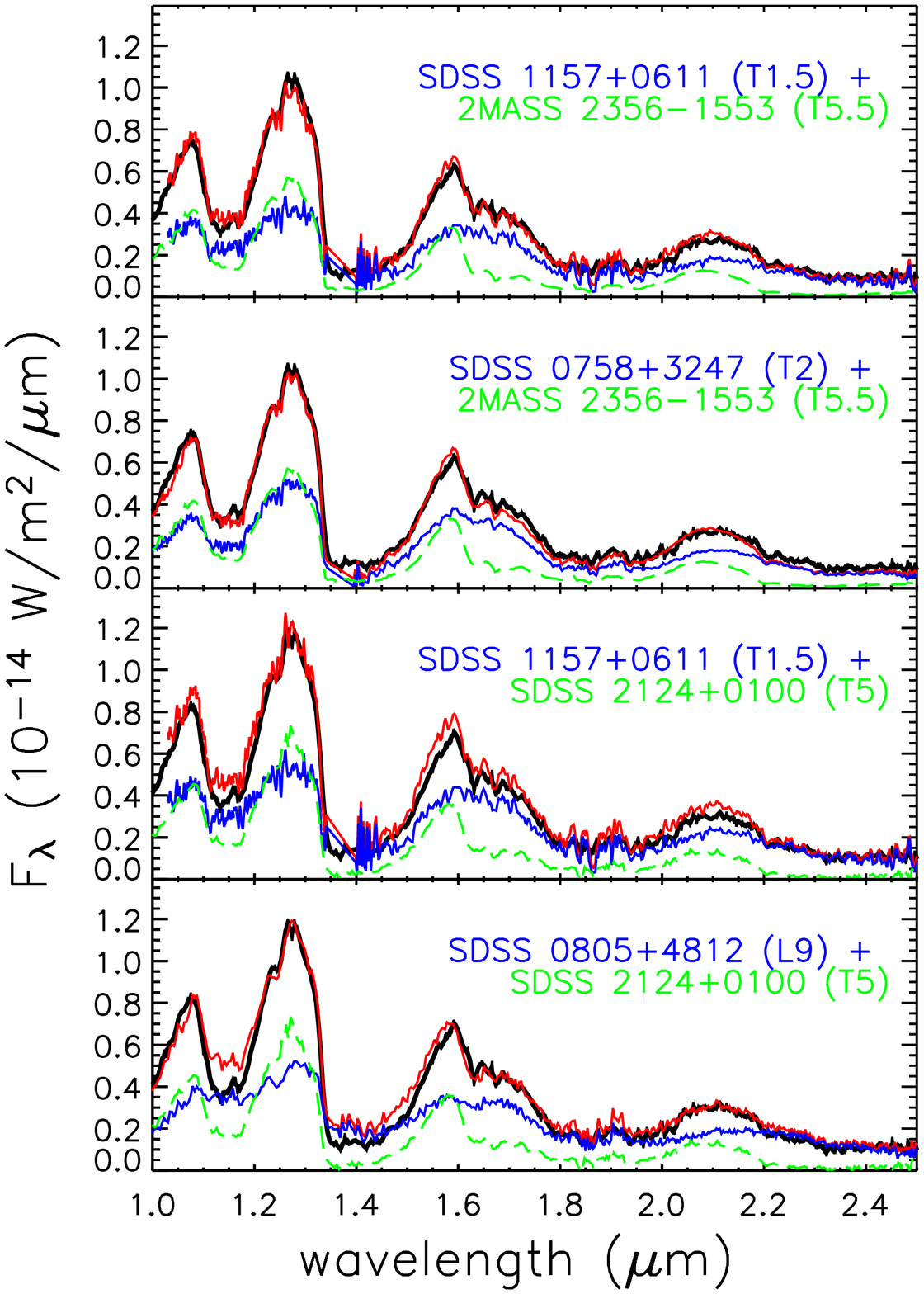}}
\vskip 3ex
\caption{\normalsize Results from modeling the integrated-light near-IR
  spectrum of \sdssbin\ (black lines; \citealp{chiu05}) as the sum of
  two T~dwarfs (red lines): one early-T template (blue solid lines) and
  one late-T template (green dashed lines).  The template objects are
  chosen to have similar IR colors as the resolved components of
  \sdssbin{AB}, and the observed $J$-band flux ratio is used in scaling
  and combining the template spectra.  The only free parameter in the
  fitting is the distance to \sdssbin{AB}.  (See \S~3 for details.)  The
  agreement between the observed spectrum and the modeled blends is
  good, especially for the topmost panel, meaning that the spectral
  types inferred for \sdssbin{A} and~B from the near-IR colors are
  consistent with the integrated-light spectrum.  The bottommost panel
  shows the match attempted using a blue late-L dwarf as component A;
  the depth of the $J$-band water absorption is not well-matched.  The
  template spectra are from \citet{chiu05}, \citet{2004AJ....127.3553K},
  and \citet{burg01}.}
\end{figure}

\begin{figure}
\vskip -2in
\hskip 0.15in
\centerline{\includegraphics[width=4in,angle=90]{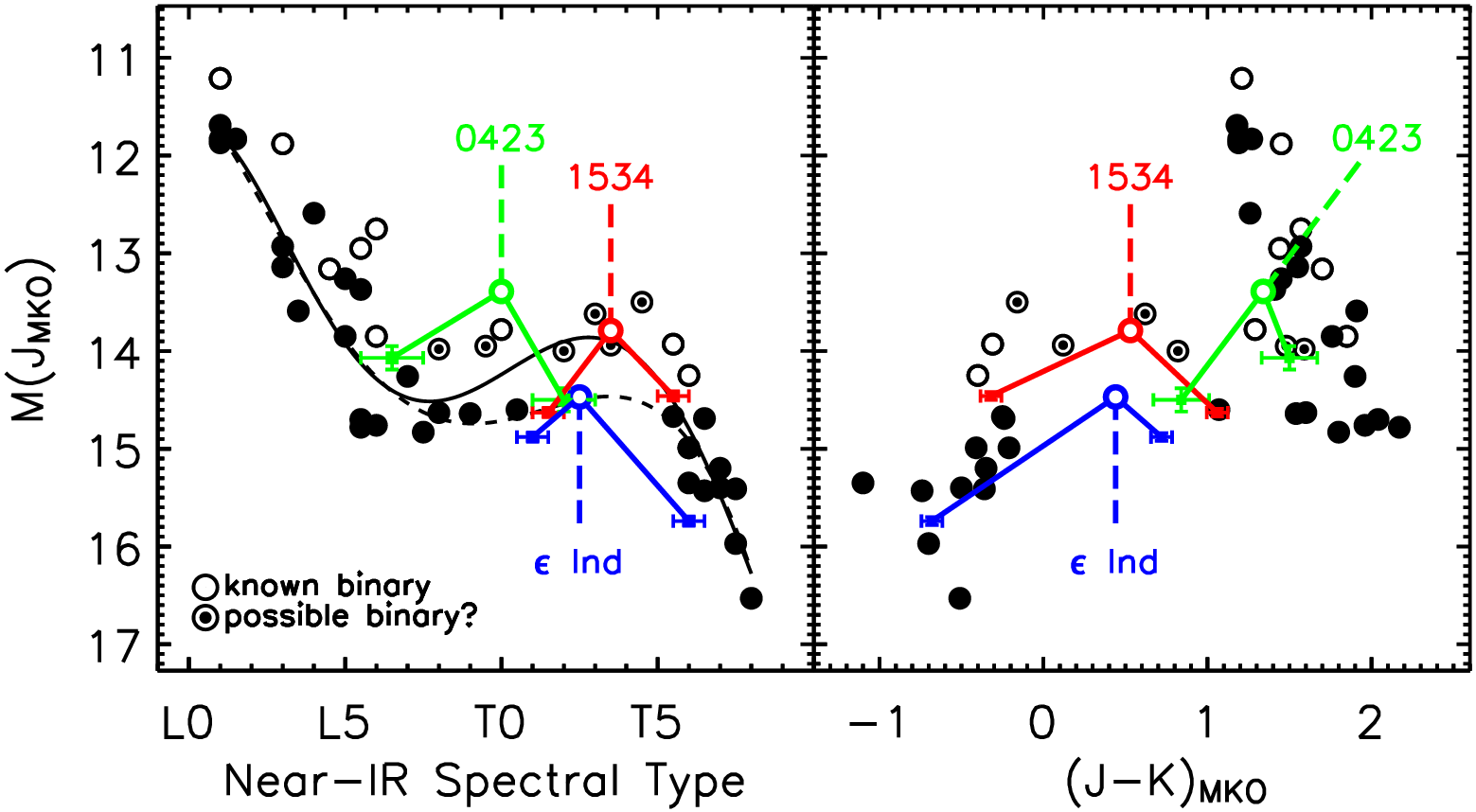}}
\caption{\normalsize $J$-band absolute magnitude as a function of
  near-IR spectral type and $J-K$ color, based on data tabulated by
  \citet{2004AJ....127.3553K} and excluding sources with S/N$<$4
  parallax measurements (SDSS~J1435$-$0043 [T6], SDSS~J0207+0000 [T4.5],
  and SDSS~J0837$-$0000 [T0.5]).  Spectral types are based on the
  \citet{geb01} scheme for the L~dwarfs and the
  \citet{2005astro.ph.10090B} scheme for the T~dwarfs.  The
  integrated-light properties of the binary T~dwarfs \sdssbin{AB},
  \eInd~Bab, and SDSS~J0423$-$0414AB are plotted as colored open
  circles, and the colored filled squares show the individual
  components.  (For the latter two objects, see \S~4 for a description
  of their resolved measurements.)  The distance to \sdssbin{AB} is
  based on modeling the integrated-light spectrum as the blend of two
  components (see \S~3), and thus is more uncertain than the
  trigonometric distances for \eInd~{Bab} and SDSS~J0423$-$0414AB.
  Errors in $M(J)$ are comparable to the symbol size. On the left, the
  solid line shows a 5th-order polynomial fit for $M(J)$ versus near-IR
  spectral type, excluding known binaries.  To illustrate the potential
  significance of binarity, the dashed line shows the same order fit,
  excluding both known and possible binaries.  Possible binaries were
  selected based on their relatively high \Teff\ compared to objects of
  similar spectral type in the \citet{gol04} measurements (see \S~4.2);
  these are SDSS~J0032+1410 (L8), 2MASS~J0328+2302 (L9.5),
  SDSS~J1254$-$0122 (T2), SDSS~J1021$-$0304 (T3), SDSS~J1750+1759
  (T3.5), and 2MASS~J0559$-$1404 (T4.5).  Table~3 provides the
  coefficients of the polynomial fits.}
\end{figure}

\begin{figure}
\hskip 0.15in
\centerline{\includegraphics[width=5in,angle=0]{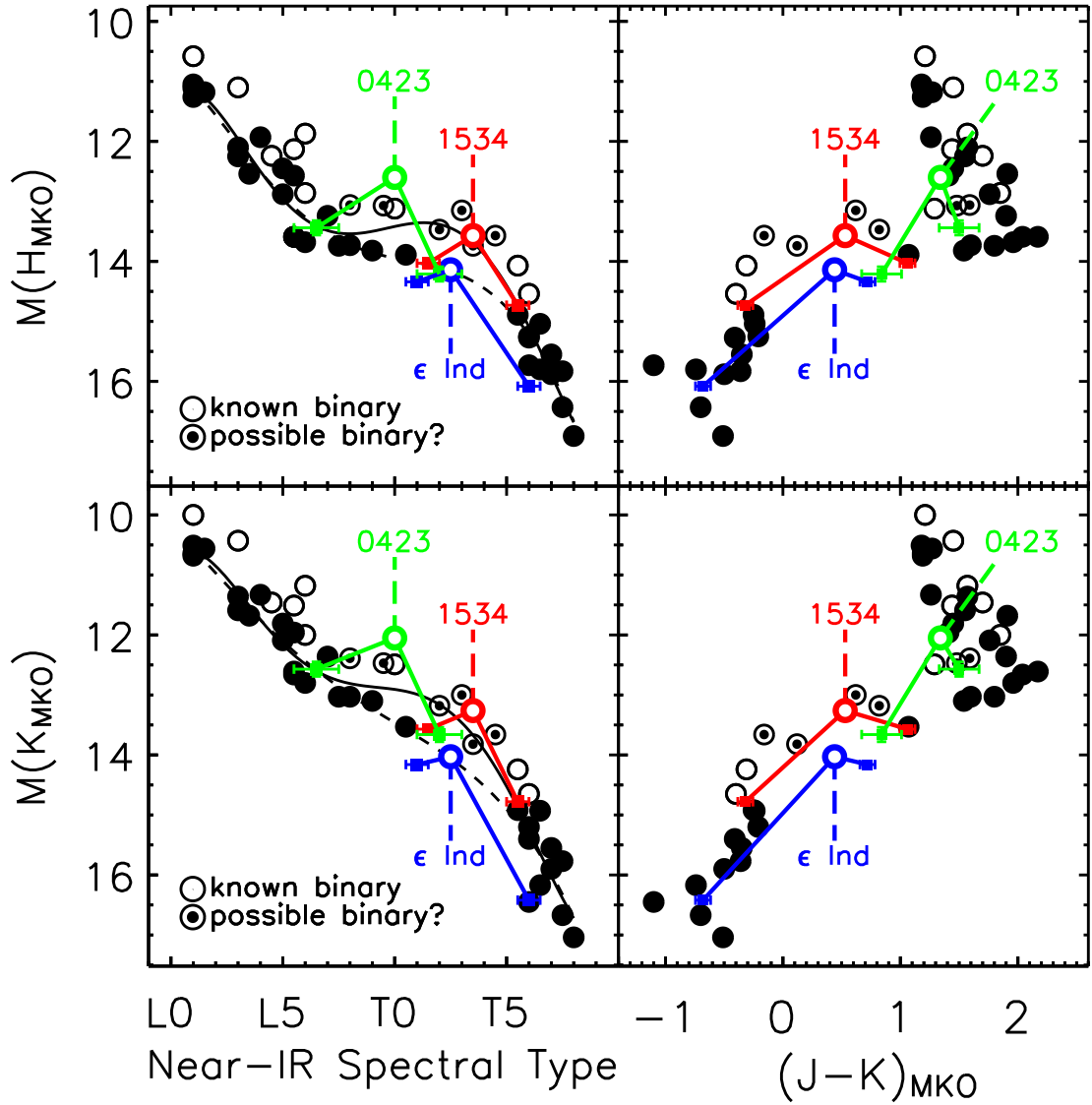}}
\caption{\normalsize $H$ and $K$-band absolute magnitudes as a function
  of near-IR spectral type and $J-K$ color.  See Figure~4 for a
  description of the content.}
\end{figure}





\clearpage

\begin{deluxetable}{lc}
\tablecaption{\sdssbin{AB} Binary Properties\tablenotemark{a} \label{sample}}
\tablewidth{0pt}
\tablehead{
  \colhead{Property} &
  \colhead{Measurement}
}

\startdata
Separation (mas)                            &  110 $\pm$ 5\phn     \\
Position angle (deg)                        &  287.6 $\pm$ 1.3\phn \\
$\Delta{J}$ (mags)                          &  $-$0.17 $\pm$ 0.04  \\
$\Delta{H}$ (mags)                          &  \phs0.70 $\pm$ 0.04 \\
$\Delta{\Kp}$ (mags)                        &  \phs1.07 $\pm$ 0.05 \\
Proper motion: $\mu_\alpha$, $\mu_\delta$ (mas\ yr\perone) & $-$142 $\pm$ 14, $-$57 $\pm$ 10 \\
Mass ratio $M_B/M_A$                        &  0.80 $\pm$ 0.05 \\
Luminosity ratio $\log(L_A/L_B)$ (dex)      &  0.3 $\pm$ 0.1 \\
Surface gravity ratio $\log(g_A/g_B)$ (dex) &  0.10 $\pm$ 0.05 \\
Estimated distance (pc)                     &  $\approx$36  \\
Total mass (\Mjup): 0.5, 1.0, 5.0 Gyr       &  60, 85, 145 \\
Estimated orbital period (yr)               &  22--33 \\
\enddata

\tablenotetext{a}{All photometry on the MKO system.}

\end{deluxetable}

\clearpage

\begin{deluxetable}{lcc}
\tablecaption{Resolved Properties of \sdssbin{AB}\tablenotemark{a}}
\tablewidth{0pt}
\tablehead{
  \colhead{Property} &
  \colhead{SDSS~J1534+1615A} &
  \colhead{SDSS~J1534+1615B}
}

\startdata
$J$ (mags)         &  17.43 $\pm$ 0.04  &    17.26 $\pm$ 0.04  \\
$H$ (mags)         &  16.83 $\pm$ 0.04  &    17.53 $\pm$ 0.04  \\
$K$ (mags)         &  16.37 $\pm$ 0.06  &    17.58 $\pm$ 0.06  \\
$J-H$ (mags)       &   0.60 $\pm$ 0.06  &  $-$0.27 $\pm$ 0.06  \\
$H-K$ (mags)       &   0.46 $\pm$ 0.07  &  $-$0.05 $\pm$ 0.07  \\
$J-K$ (mags)       &   1.06 $\pm$ 0.07  &  $-$0.32 $\pm$ 0.07  \\

Estimated spectral type\tablenotemark{b} &  T1.5 $\pm$ 0.5  &  T5.5 $\pm$ 0.5 \\
\enddata

\tablenotetext{a}{All photometry on the MKO photometric system.}
\tablenotetext{b}{Based on the \citet{2005astro.ph.10090B} classification scheme.}

\end{deluxetable}

\clearpage

\begin{deluxetable}{lcccccccc}
\tablecaption{Coefficients of Polynomial Fits to Absolute Magnitudes versus
  Near-IR Spectral Type}
\tabletypesize{\small}
\tablewidth{0pt}
\tablehead{
  \colhead{Magnitude} &
  \colhead{$c_0$} &
  \colhead{$c_1$} &
  \colhead{$c_2$} &
  \colhead{$c_3$} &
  \colhead{$c_4$} &
  \colhead{$c_5$} &
  \colhead{RMS} 
}

\startdata
\cutinhead{Excluding Known Binaries}
$M(J_{MKO})$ & 11.746 & $-$2.259e$-$1 & 3.229e$-$1 & $-$5.155e$-$2 & 2.966e$-$3 & $-$5.648e$-$5 & 0.39 \\
$M(H_{MKO})$ & 11.263 & $-$4.164e$-$1 & 3.565e$-$1 & $-$5.610e$-$2 & 3.349e$-$3 & $-$6.720e$-$5 & 0.35 \\
$M(K_{MKO})$ & 10.731 & $-$3.964e$-$1 & 3.150e$-$1 & $-$4.912e$-$2 & 2.994e$-$3 & $-$6.179e$-$5 & 0.39 \\

\cutinhead{Excluding Known and Possible Binaries}
$M(J_{MKO})$ & 11.359 & 3.174e$-$1 & 1.102e$-$1 & $-$1.877e$-$2 & 9.169e$-$4 & $-$1.233e$-$5 & 0.37 \\
$M(H_{MKO})$ & 10.767 & 2.744e$-$1 & 8.955e$-$2 & $-$1.552e$-$2 & 8.319e$-$4 & $-$1.315e$-$5 & 0.30 \\
$M(K_{MKO})$ & 10.182 & 3.688e$-$1 & 2.028e$-$2 & $-$4.488e$-$3 & 2.301e$-$4 & $-$2.466e$-$6 & 0.34 \\

\enddata

\tablecomments{These are the coefficients of the 5th-order polynomial
  fits plotted in Figures~4 and~5 for the IR absolute magnitudes as a
  function of near-IR spectral type.  The fits are defined as 
$$Magnitude = \sum_{i=0}^5 c_i \times (SpT)^{i}$$ where the numerical
  spectral type is defined as $SpT=1$ for L1, $SpT=2$ for L2, $SpT=10$
  for T0, etc.  The L~dwarfs are classified on the \citet{geb01} scheme
  and the T~dwarfs on the \citet{2005astro.ph.10090B} scheme.  The fits
  are applicable from L1 to T8.  Possible binaries were selected based
  on their relatively high \Teff\ compared to objects of similar
  spectral type --- see \S~4.2 and the caption of Figure 4.  The last
  column gives the standard deviation about the fit in magnitudes.}

\end{deluxetable}


\begin{thebibliography}{66}
\expandafter\ifx\csname natexlab\endcsname\relax\def\natexlab#1{#1}\fi

\bibitem[{Ackerman \& {Marley}(2001)}]{2001ApJ...556..872A}
Ackerman, A.~S. \& {Marley}, M.~S. 2001, \apj, 556, 872

\bibitem[{Baraffe {et~al.}(2003)Baraffe, {Chabrier}, {Barman}, {Allard}, \&
  {Hauschildt}}]{2003A&A...402..701B}
Baraffe, I., {Chabrier}, G., {Barman}, T.~S., {Allard}, F., \& {Hauschildt},
  P.~H. 2003, \aap, 402, 701

\bibitem[{Becklin \& {Zuckerman}(1988)}]{1988Natur.336..656B}
Becklin, E.~E. \& {Zuckerman}, B. 1988, \nat, 336, 656

\bibitem[{Bouchez {et~al.}(2004)}]{2004SPIE.5490..321B}
Bouchez, A.~H. {et~al.} 2004, in Advancements in Adaptive Optics. Edited by
  Domenico B. Calia, Brent L. Ellerbroek, and Roberto Ragazzoni. Proceedings of
  the SPIE., Vol. 5490, 321--330

\bibitem[{Bouy {et~al.}(2003)Bouy, {Brandner}, {Mart{\'{\i}}n}, {Delfosse},
  {Allard}, \& {Basri}}]{2003AJ....126.1526B}
Bouy, H., {Brandner}, W., {Mart{\'{\i}}n}, E.~L., {Delfosse}, X., {Allard}, F.,
  \& {Basri}, G. 2003, \aj, 126, 1526

\bibitem[{Burgasser {et~al.}(2005{\natexlab{a}})Burgasser, {Geballe},
  {Leggett}, {Kirkpatrick}, \& {Golimowski}}]{2005astro.ph.10090B}
Burgasser, A.~J., {Geballe}, T.~R., {Leggett}, S.~K., {Kirkpatrick}, J.~D., \&
  {Golimowski}, D.~A. 2005{\natexlab{a}}, ArXiv Astrophysics e-prints

\bibitem[{Burgasser {et~al.}(2003{\natexlab{a}})Burgasser, {Kirkpatrick},
  {Liebert}, \& {Burrows}}]{2003ApJ...594..510B}
Burgasser, A.~J., {Kirkpatrick}, J.~D., {Liebert}, J., \& {Burrows}, A.
  2003{\natexlab{a}}, \apj, 594, 510

\bibitem[{Burgasser {et~al.}(2005{\natexlab{b}})Burgasser, {Kirkpatrick}, \&
  {Lowrance}}]{2005AJ....129.2849B}
Burgasser, A.~J., {Kirkpatrick}, J.~D., \& {Lowrance}, P.~J.
  2005{\natexlab{b}}, \aj, 129, 2849

\bibitem[{Burgasser {et~al.}(2003{\natexlab{b}})Burgasser, {Kirkpatrick},
  {Reid}, {Brown}, {Miskey}, \& {Gizis}}]{2003ApJ...586..512B}
Burgasser, A.~J., {Kirkpatrick}, J.~D., {Reid}, I.~N., {Brown}, M.~E.,
  {Miskey}, C.~L., \& {Gizis}, J.~E. 2003{\natexlab{b}}, \apj, 586, 512

\bibitem[{Burgasser {et~al.}(2002{\natexlab{a}})Burgasser, {Marley},
  {Ackerman}, {Saumon}, {Lodders}, {Dahn}, {Harris}, \&
  {Kirkpatrick}}]{2002ApJ...571L.151B}
Burgasser, A.~J., {Marley}, M.~S., {Ackerman}, A.~S., {Saumon}, D., {Lodders},
  K., {Dahn}, C.~C., {Harris}, H.~C., \& {Kirkpatrick}, J.~D.
  2002{\natexlab{a}}, \apjl, 571, L151

\bibitem[{Burgasser {et~al.}(2006)Burgasser, {Reid}, {Leggett}, {Kirkpatrick},
  {Liebert}, \& {Burrows}}]{2005astro.ph.10580B}
Burgasser, A.~J., {Reid}, I.~N., {Leggett}, S.~K., {Kirkpatrick}, J.~D.,
  {Liebert}, J., \& {Burrows}, A. 2006, \apjl, in press
  (arXiv:astro-ph/0510580)

\bibitem[{Burgasser {et~al.}(2000)}]{2000ApJ...531L..57B}
Burgasser, A.~J. {et~al.} 2000, \apjl, 531, L57

\bibitem[{Burgasser {et~al.}(2002{\natexlab{b}})}]{burg01}
---. 2002{\natexlab{b}}, \apj, 564, 421

\bibitem[{Burrows {et~al.}(2001)Burrows, {Hubbard}, {Lunine}, \&
  {Liebert}}]{bur01}
Burrows, A., {Hubbard}, W.~B., {Lunine}, J.~I., \& {Liebert}, J. 2001, Reviews
  of Modern Physics, 73, 719

\bibitem[{Burrows {et~al.}(2006)Burrows, {Sudarsky}, \&
  {Hubeny}}]{2005astro.ph..9066B}
Burrows, A., {Sudarsky}, D., \& {Hubeny}, I. 2006, \apj, in press
  (astro-ph/0509066)

\bibitem[{Chiu {et~al.}(2006)}]{chiu05}
Chiu, K. {et~al.} 2006, \apj, in press (astro-ph/0601089)

\bibitem[{Cruz {et~al.}(2004)Cruz, {Burgasser}, {Reid}, \&
  {Liebert}}]{2004ApJ...604L..61C}
Cruz, K.~L., {Burgasser}, A.~J., {Reid}, I.~N., \& {Liebert}, J. 2004, \apjl,
  604, L61

\bibitem[{Cushing {et~al.}(2005)Cushing, {Rayner}, \&
  {Vacca}}]{2005ApJ...623.1115C}
Cushing, M.~C., {Rayner}, J.~T., \& {Vacca}, W.~D. 2005, \apj, 623, 1115

\bibitem[{Dahn {et~al.}(2002)}]{2002AJ....124.1170D}
Dahn, C.~C. {et~al.} 2002, \aj, 124, 1170

\bibitem[{Diolaiti {et~al.}(2000)Diolaiti, {Bendinelli}, {Bonaccini}, {Close},
  {Currie}, \& {Parmeggiani}}]{2000A&AS..147..335D}
Diolaiti, E., {Bendinelli}, O., {Bonaccini}, D., {Close}, L., {Currie}, D., \&
  {Parmeggiani}, G. 2000, \aaps, 147, 335

\bibitem[{Enoch {et~al.}(2003)Enoch, {Brown}, \&
  {Burgasser}}]{2003AJ....126.1006E}
Enoch, M.~L., {Brown}, M.~E., \& {Burgasser}, A.~J. 2003, \aj, 126, 1006

\bibitem[{Fischer \& {Marcy}(1992)}]{1992ApJ...396..178F}
Fischer, D.~A. \& {Marcy}, G.~W. 1992, \apj, 396, 178

\bibitem[{Freed {et~al.}(2003)Freed, {Close}, \&
  {Siegler}}]{2003ApJ...584..453F}
Freed, M., {Close}, L.~M., \& {Siegler}, N. 2003, \apj, 584, 453

\bibitem[{Geballe {et~al.}(2002)}]{geb01}
Geballe, T. {et~al.} 2002, \apj, 564, 466

\bibitem[{Gelino {et~al.}(2004)Gelino, {Kirkpatrick}, \&
  {Burgasser}}]{2004AAS...205.1113G}
Gelino, C.~R., {Kirkpatrick}, J.~D., \& {Burgasser}, A.~J. 2004, \baas, 205

\bibitem[{Gizis {et~al.}(2003)Gizis, {Reid}, {Knapp}, {Liebert}, {Kirkpatrick},
  {Koerner}, \& {Burgasser}}]{2003AJ....125.3302G}
Gizis, J.~E., {Reid}, I.~N., {Knapp}, G.~R., {Liebert}, J., {Kirkpatrick},
  J.~D., {Koerner}, D.~W., \& {Burgasser}, A.~J. 2003, \aj, 125, 3302

\bibitem[{Golimowski {et~al.}(2004{\natexlab{a}})}]{gol04}
Golimowski, D.~A. {et~al.} 2004{\natexlab{a}}, \aj, 127, 3516

\bibitem[{Golimowski {et~al.}(2004{\natexlab{b}})}]{2004AJ....128.1733G}
---. 2004{\natexlab{b}}, \aj, 128, 1733

\bibitem[{Kaiser {et~al.}(2002)}]{2002SPIE.4836..154K}
Kaiser, N. {et~al.} 2002, in Survey and Other Telescope Technologies and
  Discoveries. Edited by Tyson, J. Anthony; Wolff, Sidney. Proceedings of the
  SPIE, Volume 4836, pp. 154-164 (2002)., 154--164

\bibitem[{Kirkpatrick(2003)}]{2003IAUS..211..189K}
Kirkpatrick, J.~D. 2003, in Proceedings of IAU Symposium 211: Brown Dwarfs, ed.
  E.~Martin, 189

\bibitem[{Kirkpatrick(2005)}]{kirk05}
Kirkpatrick, J.~D. 2005, \araa, 43, 195

\bibitem[{Kirkpatrick {et~al.}(2001)Kirkpatrick, {Dahn}, {Monet}, {Reid},
  {Gizis}, {Liebert}, \& {Burgasser}}]{2001AJ....121.3235K}
Kirkpatrick, J.~D., {Dahn}, C.~C., {Monet}, D.~G., {Reid}, I.~N., {Gizis},
  J.~E., {Liebert}, J., \& {Burgasser}, A.~J. 2001, \aj, 121, 3235

\bibitem[{Kirkpatrick {et~al.}(2000)Kirkpatrick, {Reid}, {Liebert}, {Gizis},
  {Burgasser}, {Monet}, {Dahn}, {Nelson}, \& {Williams}}]{2000AJ....120..447K}
Kirkpatrick, J.~D., {Reid}, I.~N., {Liebert}, J., {Gizis}, J.~E., {Burgasser},
  A.~J., {Monet}, D.~G., {Dahn}, C.~C., {Nelson}, B., \& {Williams}, R.~J.
  2000, \aj, 120, 447

\bibitem[{Knapp {et~al.}(2004)}]{2004AJ....127.3553K}
Knapp, G.~R. {et~al.} 2004, \aj, 127, 3553

\bibitem[{Koen {et~al.}(2004)Koen, {Matsunaga}, \&
  {Menzies}}]{2004MNRAS.354..466K}
Koen, C., {Matsunaga}, N., \& {Menzies}, J. 2004, \mnras, 354, 466

\bibitem[{Koen {et~al.}(2005)Koen, {Tanab{\'e}}, {Tamura}, \&
  {Kusakabe}}]{2005MNRAS.362..727K}
Koen, C., {Tanab{\'e}}, T., {Tamura}, M., \& {Kusakabe}, N. 2005, \mnras, 362,
  727

\bibitem[{Koerner {et~al.}(1999)Koerner, {Kirkpatrick}, {McElwain}, \&
  {Bonaventura}}]{1999ApJ...526L..25K}
Koerner, D.~W., {Kirkpatrick}, J.~D., {McElwain}, M.~W., \& {Bonaventura},
  N.~R. 1999, \apjl, 526, L25

\bibitem[{{Leggett} {et~al.}(2004){Leggett}, {Allard}, {Burgasser}, {Jones},
  {Marley}, \& {Tsuji}}]{2004astro.ph..9389L}
{Leggett}, S.~K., {Allard}, F., {Burgasser}, A.~J., {Jones}, H.~R.~A.,
  {Marley}, M.~S., \& {Tsuji}, T. 2004, ArXiv Astrophysics e-prints

\bibitem[{Leggett {et~al.}(2002)}]{leg01}
Leggett, S.~K. {et~al.} 2002, \apj, 564, 452

\bibitem[{Liu {et~al.}(2002)Liu, {Fischer}, {Graham}, {Lloyd}, {Marcy}, \&
  {Butler}}]{2001astro.ph.12407L}
Liu, M.~C., {Fischer}, D.~A., {Graham}, J.~R., {Lloyd}, J.~P., {Marcy}, G.~W.,
  \& {Butler}, R.~P. 2002, \apj, 571, 519

\bibitem[{Liu \& {Leggett}(2005)}]{2005astro.ph..8082L}
Liu, M.~C. \& {Leggett}, S.~K. 2005, \apj, 634, 616

\bibitem[{Marley {et~al.}(2002)Marley, {Seager}, {Saumon}, {Lodders},
  {Ackerman}, {Freedman}, \& {Fan}}]{2002ApJ...568..335M}
Marley, M.~S., {Seager}, S., {Saumon}, D., {Lodders}, K., {Ackerman}, A.~S.,
  {Freedman}, R.~S., \& {Fan}, X. 2002, \apj, 568, 335

\bibitem[{Mart{\'{\i}}n {et~al.}(1999)Mart{\'{\i}}n, {Brandner}, \&
  {Basri}}]{1999Sci...283.1718M}
Mart{\'{\i}}n, E.~L., {Brandner}, W., \& {Basri}, G. 1999, Science, 283, 1718

\bibitem[{Maxted \& {Jeffries}(2005)}]{2005MNRAS.362L..45M}
Maxted, P.~F.~L. \& {Jeffries}, R.~D. 2005, \mnras, 362, L45

\bibitem[{McCaughrean {et~al.}(2004)McCaughrean, {Close}, {Scholz}, {Lenzen},
  {Biller}, {Brandner}, {Hartung}, \& {Lodieu}}]{2004A&A...413.1029M}
McCaughrean, M.~J., {Close}, L.~M., {Scholz}, R.-D., {Lenzen}, R., {Biller},
  B., {Brandner}, W., {Hartung}, M., \& {Lodieu}, N. 2004, \aap, 413, 1029

\bibitem[{Metchev \& {Hillenbrand}(2004)}]{2004ApJ...617.1330M}
Metchev, S.~A. \& {Hillenbrand}, L.~A. 2004, \apj, 617, 1330

\bibitem[{Monet {et~al.}(2003)}]{2003AJ....125..984M}
Monet, D.~G. {et~al.} 2003, \aj, 125, 984

\bibitem[{Nakajima {et~al.}(1995)Nakajima, {Oppenheimer}, {Kulkarni},
  {Golimowski}, {Matthews}, \& {Durrance}}]{1995Natur.378..463N}
Nakajima, T., {Oppenheimer}, B.~R., {Kulkarni}, S.~R., {Golimowski}, D.~A.,
  {Matthews}, K., \& {Durrance}, S.~T. 1995, \nat, 378, 463

\bibitem[{Nakajima {et~al.}(2001)Nakajima, {Tsuji}, \&
  {Yanagisawa}}]{2001ApJ...561L.119N}
Nakajima, T., {Tsuji}, T., \& {Yanagisawa}, K. 2001, \apjl, 561, L119

\bibitem[{Nakajima {et~al.}(2004)Nakajima, {Tsuji}, \&
  {Yanagisawa}}]{2004ApJ...607..499N}
---. 2004, \apj, 607, 499

\bibitem[{Perryman {et~al.}(1997)}]{1997A&A...323L..49P}
Perryman, M.~A.~C. {et~al.} 1997, \aap, 323, L49

\bibitem[{Potter {et~al.}(2002)Potter, {Mart{\' i}n}, {Cushing}, {Baudoz},
  {Brandner}, {Guyon}, \& {Neuh{\" a}user}}]{2002ApJ...567L.133P}
Potter, D., {Mart{\' i}n}, E.~L., {Cushing}, M.~C., {Baudoz}, P., {Brandner},
  W., {Guyon}, O., \& {Neuh{\" a}user}, R. 2002, \apjl, 567, L133

\bibitem[{Rebolo {et~al.}(1998)Rebolo, {Zapatero Osorio}, {Madruga}, {Bejar},
  {Arribas}, \& {Licandro}}]{1998Sci...282.1309R}
Rebolo, R., {Zapatero Osorio}, M.~R., {Madruga}, S., {Bejar}, V.~J.~S.,
  {Arribas}, S., \& {Licandro}, J. 1998, Science, 282, 1309

\bibitem[{Reid {et~al.}(2001)Reid, {Gizis}, {Kirkpatrick}, \&
  {Koerner}}]{2001AJ....121..489R}
Reid, I.~N., {Gizis}, J.~E., {Kirkpatrick}, J.~D., \& {Koerner}, D.~W. 2001,
  \aj, 121, 489

\bibitem[{Reid {et~al.}(2006)Reid, {Lewitus}, {Burgasser}, \&
  {Cruz}}]{2006ApJ...639.1114R}
Reid, I.~N., {Lewitus}, E., {Burgasser}, A.~J., \& {Cruz}, K.~L. 2006, \apj,
  639, 1114

\bibitem[{Scholz {et~al.}(2003)Scholz, {McCaughrean}, {Lodieu}, \&
  {Kuhlbrodt}}]{2003A&A...398L..29S}
Scholz, R.-D., {McCaughrean}, M.~J., {Lodieu}, N., \& {Kuhlbrodt}, B. 2003,
  \aap, 398, L29

\bibitem[{Simons \& {Tokunaga}(2002)}]{mkofilters1}
Simons, D.~A. \& {Tokunaga}, A. 2002, \pasp, 114, 169

\bibitem[{Thorstensen \& {Kirkpatrick}(2003)}]{2003PASP..115.1207T}
Thorstensen, J.~R. \& {Kirkpatrick}, J.~D. 2003, \pasp, 115, 1207

\bibitem[{Tinney {et~al.}(2003)Tinney, {Burgasser}, \&
  {Kirkpatrick}}]{2003AJ....126..975T}
Tinney, C.~G., {Burgasser}, A.~J., \& {Kirkpatrick}, J.~D. 2003, \aj, 126, 975

\bibitem[{Tokunaga {et~al.}(2002)Tokunaga, {Simons}, \& {Vacca}}]{mkofilters2}
Tokunaga, A.~T., {Simons}, D.~A., \& {Vacca}, W.~D. 2002, \pasp, 114, 180

\bibitem[{Torres(1999)}]{1999PASP..111..169T}
Torres, G. 1999, \pasp, 111, 169

\bibitem[{{Tsuji}(2002)}]{2002ApJ...575..264T}
{Tsuji}, T. 2002, \apj, 575, 264

\bibitem[{{Tsuji}(2005)}]{2005ApJ...621.1033T}
---. 2005, \apj, 621, 1033

\bibitem[{{Tsuji} \& {Nakajima}(2003)}]{2003ApJ...585L.151T}
{Tsuji}, T. \& {Nakajima}, T. 2003, \apjl, 585, L151

\bibitem[{Vrba {et~al.}(2004)}]{2004AJ....127.2948V}
Vrba, F.~J. {et~al.} 2004, \aj, 127, 2948

\bibitem[{Wizinowich {et~al.}(2004)}]{2004SPIE.5490....1W}
Wizinowich, P.~L. {et~al.} 2004, in Advancements in Adaptive Optics. Edited by
  Domenico B. Calia, Brent L. Ellerbroek, and Roberto Ragazzoni. Proceedings of
  the SPIE., Vol. 5490, 1--11

\end{thebibliography}
\end{document}